%% file: neurips_2026.tex
\documentclass{article}

    \PassOptionsToPackage{numbers, compress}{natbib}

\usepackage[position, final]{neurips_2026}

\usepackage[utf8]{inputenc}
\usepackage[T1]{fontenc}
\usepackage{hyperref}
\usepackage{url}
\usepackage{booktabs}
\usepackage{amsfonts}
\usepackage{nicefrac}
\usepackage{microtype}
\usepackage{xcolor}

\usepackage{amsmath}
\usepackage{algorithm}
\usepackage{algpseudocode}
\usepackage{wrapfig}
\usepackage{subcaption}
\usepackage{graphicx}
\usepackage{multirow}
\usepackage{amsthm}
\usepackage[most]{tcolorbox}
\tcbuselibrary{theorems}
\usepackage[capitalize]{cleveref}
\usepackage{enumitem}

\setlist{topsep=2pt, itemsep=1pt, parsep=0pt}

\makeatletter
\renewcommand{\paragraph}{%
  \@startsection{paragraph}{4}{\z@}%
                {0pt \@plus 0.05\baselineskip \@minus 0.05\baselineskip}%
                {-1em}%
                {\normalsize\bf}%
}
\renewcommand{\section}{\@startsection{section}{1}{\z@}%
  {-1.4ex \@plus -0.4ex \@minus -0.2ex}%
  { 0.8ex \@plus  0.2ex}%
  {\large\bf\raggedright}}
\renewcommand{\subsection}{\@startsection{subsection}{2}{\z@}%
  {-1.2ex \@plus -0.4ex \@minus -0.2ex}%
  { 0.4ex \@plus  0.2ex}%
  {\normalsize\bf\raggedright}}
\renewcommand{\subsubsection}{\@startsection{subsubsection}{3}{\z@}%
  {-1.0ex \@plus -0.4ex \@minus -0.2ex}%
  { 0.3ex \@plus  0.2ex}%
  {\normalsize\bf\raggedright}}
\makeatother

\title{Epistemic Infrastructures of Science in AI Era Should Rebalance Costs of Generation and Verification}
\newcommand{\blueprintRepoUrl}{\url{https://github.com/jiaqima/blueprint}}

\author{%
  Jiaqi W. Ma \\
  University of Illinois Urbana-Champaign \\
  \texttt{jiaqima@illinois.edu} \\
}

\begin{document}

\maketitle

\begin{abstract}
    \input{sections/0-abstract}
\end{abstract}

\input{sections/1-intro}
\input{sections/2-problem}
\input{sections/3-cause}
\input{sections/4-solution}
\input{sections/5-related}
\input{sections/6-alternatives}
\input{sections/7-discussion}
\input{sections/8-conclusion}

\begin{ack}
JM would like to acknowledge support by NSF CAREER Award No.~2542856.
\end{ack}

\bibliographystyle{plainnat}
\bibliography{references}

\appendix

\input{sections/appendix}

\end{document}

%% file: sections/0-abstract.tex
AI systems can now cheaply generate plausible scientific artifacts such as papers, reviews, and surveys. This creates a risk of \emph{epistemic pollution} in our scientific systems, where unreliable but plausible-looking artifacts can accumulate faster than the system can filter them out. The problem is structural: the epistemic infrastructure of science was calibrated to a world where producing a plausible artifact required substantial expertise, labor, and time, so generation cost itself served as a rough filter; AI weakens that filter without comparably lowering verification cost. This position paper argues that \textbf{AI-era science should rebalance the costs of generating and verifying scientific artifacts by redesigning epistemic infrastructure at the level of the artifact}. The current paper-centered system makes verification expensive: papers compress long-context scientific logic into prose, forcing reviewers, human or AI, to reconstruct underlying argument structure before they can evaluate it. As one step in this direction, we propose \textbf{blueprints}: structured, decomposed research artifacts that represent claims, evidence, assumptions, and definitions as typed graph components. Blueprints complement rather than replace papers---the paper remains the narrative interface, the blueprint becomes the verification interface---and are designed to trade an upfront generation cost for cheaper, more local, more distributed verification downstream. We also instantiate the proposal in a proof-of-concept prototype.

%% file: sections/1-intro.tex
\section{Introduction}\label{sec:intro}

AI is reshaping the cost structure of scientific work: papers, reviews, hypotheses, code, and literature surveys can now be generated at low cost~\citep{Lu2024TheAS,tang2025ai,Yamada2025TheAS}. But science is fundamentally the production of verifiable, reusable, and trustworthy knowledge. A central bottleneck of AI-era science is therefore shifting from generating plausible research artifacts toward verifying them to maintain a reliable scientific knowledge base. The existing \emph{epistemic infrastructure} of science---the artifacts, workflows, and norms through which claims are produced, reviewed, and accepted~\citep{merton1973sociology,hicks2015bibliometrics}---was developed for a world where producing a plausible scientific artifact demanded substantial expertise, labor, and time. Historically, high generation costs served as a rough filter for unreliable artifacts. AI dramatically lowers generation costs but has not proportionally reduced verification costs.

This imbalance creates conditions for \emph{epistemic pollution}: the contamination of scientific literature by plausible yet unreliable artifacts~\citep{haider2024gpt,naddaf2025major,naddaf2026hallucinated}. Examples include AI-generated surveys flooding preprint platforms, termed a ``survey paper DDoS attack''~\citep{lin2025stop}, and AI-generated peer reviews at major venues~\citep{naddaf2025major}. Additionally, automated verification systems currently show limited effectiveness in identifying significant errors or reproducibility issues~\citep{Son2025WhenAC,hu2025repro}. We diagnose epistemic pollution as a cost-ratio problem rather than merely a bad-actor problem: better frontier AI cannot be relied on to resolve it on its own, because paper mills, citation farms, and low-effort actors can exploit any system that rewards plausible artifacts while making verification costly. \textbf{We argue that addressing this challenge calls for an engineering view of science: deliberately redesigning epistemic infrastructure to rebalance the costs of generating and verifying scientific artifacts---an intervention at the level of the artifact, the unit on which the ecosystem's filters operate.}

Verification evaluates several dimensions of scientific claims, including novelty, significance, and validity. This paper focuses on \emph{validity}---whether claims are actually supported by evidence, methods, assumptions, and reasoning---for two reasons. First, validity is often especially labor-intensive: reviewers can sometimes make preliminary judgments about novelty and significance from an abstract, a figure, and their knowledge of the field, but checking whether claims hold requires tracing extended chains of reasoning through evidence, methods, and assumptions. Second, validity is the dimension where artifact design has the clearest direct leverage, because it is fundamentally about the internal logic of an argument, and how an argument is presented can either expose or hide that logic. Papers, the current dominant form of research artifacts, often obscure logical dependencies: they compress multi-step reasoning into prose, and verifying long, complex, prose-bound logic is hard for both human reviewers and AI systems.

To mitigate the cost of verifying validity, we propose restructuring scientific artifacts from prose-first narratives into \emph{structured}, \emph{decomposed} representations of shorter local arguments. We take inspiration from \texttt{leanblueprint}\footnote{\url{https://github.com/PatrickMassot/leanblueprint}} for Lean formalization~\citep{de2015lean,yang2023leandojo}, where a long, complex theorem is decomposed into a graph of shorter, simpler claims that can be worked on locally and verified with distributed collaboration, including from non-expert mathematicians. We adapt this spirit to general research, while recognizing that science beyond mathematics admits no formal proof obligation: the goal is not deductive certification but easier inspection.

We instantiate this idea in \textbf{blueprints}, a proof-of-concept prototype for representing a research project as a typed argument graph\footnote{\blueprintRepoUrl}. A blueprint decomposes a project into entities such as \emph{claims}, \emph{evidence}, \emph{assumptions}, and \emph{definitions}, connected by relations such as \emph{supports}, \emph{expands}, and \emph{contradicts}. \emph{Blueprints complement rather than replace papers}: the paper remains the narrative interface, while the blueprint acts as the verification interface. Constructing a blueprint is not free---as in \texttt{leanblueprint}, it requires expert judgment about how to decompose an argument, and good decompositions vary across fields and standards of evidence---but the intended tradeoff is exactly the rebalancing we seek: the author records the support structure once so that reviewers, readers, AI systems, and structural tools can inspect it more locally downstream. We do not present blueprints as a new theory of argument or a new semantic-publishing ontology; instead, they adapt existing structured-argument ideas~\citep{toulmin2003uses,kunz1970issues} into an author-facing artifact that makes support structure explicit and locally inspectable in AI-era science.

This paper offers three key contributions.
\begin{enumerate}
    \item \textbf{Epistemic pollution and the generation-verification imbalance.} We argue that the cost imbalance between cheap AI generation and expensive verification creates conditions for epistemic pollution: the contamination of scientific knowledge systems by plausible but unreliable artifacts.
    \item \textbf{Validity as the verification bottleneck.} We distinguish novelty, significance, and validity, and argue that validity verification is often especially expensive because prose papers compress multi-step reasoning into narrative form.
    \item \textbf{Blueprints as cost-rebalancing artifacts.} We propose blueprints as structured, decomposed research artifacts that complement papers as their verification interface, designed to trade a generation cost of argument decomposition for cheaper downstream verification. We also instantiate the concept in a working prototype.
\end{enumerate}

Together, these contributions frame one step toward an engineering of science under AI abundance: scientific artifacts should be designed not only for effective communication, but also for robust verification.

%% file: sections/2-problem.tex
\section{Epistemic Pollution and Generation-Verification Imbalance}\label{sec:pollution}

\subsection{Epistemic Pollution}

We use \emph{epistemic pollution} to refer to the contamination of scientific literature and adjacent knowledge systems---preprint repositories, peer review records, citation graphs, benchmark leaderboards, replication corpora, and AI training data---by artifacts that look like legitimate scientific contributions but cannot be relied upon. The phrase is meant to capture more than fraud, predatory publishing, or public misinformation, each of which is a distinct and longer-standing concern. Pollution sits between these: it includes negligent and low-effort artifacts that are not strictly fraudulent, reaches into mainstream venues rather than only the predatory fringe, and degrades the infrastructure scientists rely on to do their own work rather than the channels through which science reaches the public.

The phenomenon is already visible across the surface area of the literature. Large-scale analyses now find substantial LLM-associated changes in scientific writing and production patterns~\citep{liang2025quantifying,kusumegi2025scientific}. In October 2025, arXiv's computer science category changed its moderation policy for survey articles, requiring them to have passed through a successful peer review. This change is attributed to the fact that generative AIs have made papers without new research results fast and easy to write~\citep{Jones2026LeadingPS}. Automated survey-writing systems make the same capability concrete~\citep{wang2024autosurvey}. A recent commentary describes the resulting volume pressure in starker terms, framing the volume of AI-generated surveys flooding preprint platforms as a ``survey paper DDoS attack''~\citep{lin2025stop}.

Peer review itself is under similar pressure. A study in \emph{Nature} reported that 21\% of manuscript reviews for a major AI conference were found to be generated by AI~\citep{naddaf2025major}, raising the prospect of a feedback loop in which AI-generated submissions are evaluated by AI-generated reviews~\citep{kim2025position,wei2025ai}. Adjacent infrastructure---journal AI-use policies, paper mills, authorship metadata---is similarly affected~\citep{he2026academic,spinellis2025false}; \Cref{app:pollution-evidence} collects further evidence. The authors of AI Scientist-v2 report that an end-to-end autonomous research system could author a manuscript that exceeded an ICLR workshop's average human acceptance threshold~\citep{Yamada2025TheAS,Lu2024TheAS}. This result is often read as a milestone in AI capability. We read it differently: candidate generation has become powerful enough that the surrounding verification infrastructure is now the bottleneck.

\subsection{A Shift in the Cost Ratio}

The reason epistemic pollution is difficult to contain is not simply because AI introduces unreliable artifacts into science. Unreliable artifacts existed long before AI, and the existing infrastructure of peer review, citation, and venue prestige was built to filter them. What has changed is not the presence of unreliable artifacts but the \emph{cost} of producing one---and that cost was the implicit parameter on which all of these filters depended.

Producing a plausible paper, review, survey, or benchmark report used to require time, expertise, data, analysis, and revision; authorship itself functioned as a costly signal of seriousness. AI can substantially lower much of that production cost. The cost of \emph{verifying} such an artifact has not fallen to a comparable degree. Automated verification, the natural defensive response, is itself currently weak: recent benchmarks for detecting significant errors in scientific manuscripts find recall far below what would be needed for meaningful filtering~\citep{Son2025WhenAC}, and benchmarks for automated reproducibility assessment in social science report similar gaps~\citep{hu2025repro}. AI-generated artifacts have also introduced new failure modes that verification must screen for---hallucinated citations~\citep{naddaf2026hallucinated,sakai2026hallucitation}, unfaithful summaries, code that does not match the described method, results not produced by the stated procedure. At this time, the cost ratio between generating and checking a scientific artifact is sharply imbalanced.

Under such an imbalance, the existing filters do not break on their own terms---they simply lose the cost differential that gave them their discriminating power. The diagnosis is therefore infrastructural rather than behavioral: not that authors or reviewers have stopped doing their jobs, but that the infrastructure they operate within was calibrated to a cost ratio that no longer holds.

The shift in cost ratio per artifact also escalates to problems at the ecosystem level. An ecosystem's filtering capacity is the aggregate of many per-artifact verification efforts, and it remains sustainable only while the per-artifact verification cost, multiplied by the volume of artifacts, stays within the community's collective capacity.

\subsection{Why Better AI Alone Is Not Enough}

A natural response to the foregoing is to expect that better AI systems should be enough to resolve the problem: stronger generators will produce fewer hallucinations, and stronger verifiers will catch more errors. We are skeptical of relying on this alone for two reasons.

First, the problem is structural, not only capability-bound. Even if frontier AI systems become highly reliable scientific collaborators, the infrastructure remains exploitable by anyone willing to use less reliable systems. Paper mills, citation farms, low-effort authors, fake reviewers, automated submission systems, and institutions rewarded by publication counts all benefit from artifacts that pass existing filters without improving knowledge. The infrastructure rewards plausibility; as long as plausibility is cheap to produce, the most expensive part of the system---verification---bears the load, and improvements in the best generators do little to relieve it.

Second, the asymmetry between generation and verification does not scale symmetrically with capability. A capability improvement in generation lets one actor produce many more plausible artifacts at marginal cost. A capability improvement in verification, by contrast, must be applied per artifact, and verification of a complex scientific claim does not parallelize as readily as generation of a plausible-looking one. The capability side of this asymmetry is empirically visible: models can produce expert-level outputs while failing to understand or critically interrogate those same outputs~\citep{west2024generative}, and LLMs cannot reliably identify errors in their own reasoning without external feedback~\citep{huang2024large}---improvements in generation capability do not automatically produce commensurate improvements in verification capability. Even under optimistic assumptions about verifier capability, the cost of producing and the cost of checking a scientific artifact remain on different scaling curves.

The redesign question that follows is therefore not only whether AI systems can be improved, but how the infrastructure in which AI-generated and AI-assisted artifacts circulate should be redesigned to handle the cost structure AI has produced. The next section opens up verification cost itself, asking what makes it expensive and where artifact design has the most leverage.

%% file: sections/3-cause.tex
\section{Examining the Verification Cost}\label{sec:verification}

If the central problem is a cost imbalance between cheap generation and expensive verification, the natural next question is what makes verification expensive in the first place. This section decomposes verification into its components, locates where cost concentrates, and identifies the part of verification that artifact design can most directly affect.

\subsection{What Verification Costs}\label{sec:verification:components}

A reviewer, reader, or downstream user who wants to determine whether a scientific artifact is reliable cannot do so by inspection of the artifact's surface. Plausibility is what generation produces; reliability is what verification has to check. The work involved typically includes some combination of the following: identifying the central claims the artifact makes; locating the evidence offered in support of each claim; recovering the assumptions on which the argument depends, including ones the author has not stated explicitly; checking that the claimed support relations actually hold, rather than merely being asserted; evaluating whether the methods chosen are appropriate to the question; and tracking the risks, limitations, contradictions, and open questions that qualify the central claims~\citep{bornmann2011scientific,lee2013bias}.

Two features of this list matter for the rest of the section. First, the components are not equally costly. Some can be done quickly from a skim; others require deep engagement with the argument's internal logic. Second, every component is paid \emph{per verifier}. The author bears the cost of structuring an argument once; each subsequent reader, reviewer, or AI system that needs to verify it bears a reconstruction cost in turn. This asymmetry compounds: a small reduction in per-artifact verification cost is paid forward to every downstream verifier, and a small increase compounds the same way.

Verification cost, in other words, is not a flat tax on the scientific record. It is a distribution across components, paid many times by many parties, and the question of where to intervene depends on where the cost concentrates.

\subsection{Validity as the Cost Concentration}\label{sec:verification:validity}

Scientific review is conventionally said to evaluate several dimensions, of which the most commonly cited are novelty (does this contribute something the field did not already know?), significance (does the contribution matter, and to whom?), and validity (are the claims actually supported by the evidence, methods, assumptions, and reasoning offered?)~\citep{lee2013bias,bornmann2011scientific}. These dimensions are not exhaustive---reviewers also evaluate clarity, scope, and fit~\citep{lee2013bias}---but they are the dimensions that most directly determine whether an artifact enters the trusted knowledge base.

Novelty and significance, on close inspection, are partly social judgments. Whether a result is novel depends on what the field has previously established, which a reviewer often knows independently of the artifact; whether it is significant depends on what the field cares about, which is similarly external. In many cases, a reviewer can form an initial judgment on both from an abstract, a figure, and a sense of the field, even if a final judgment may require deeper engagement. Artifact design has more limited direct leverage here: a more structured artifact does not change whether a result is genuinely new or important, only how clearly it is communicated.

Validity is different in two ways. First, it is often especially labor-intensive. Where novelty and significance can sometimes be judged initially at the surface, validity requires tracing extended chains of reasoning through evidence, methods, and assumptions---running through nearly every component listed in Section~\ref{sec:verification:components}. Second, validity is the component whose cost depends most directly on how the artifact is internally organized. A claim's validity is fundamentally a property of the argument structure connecting it to its evidence, and the artifact is what records (or fails to record) that structure. Novelty and significance live partly in the field's collective context; validity lives more directly in the artifact itself.

This is the asymmetry that motivates the rest of the paper. Verification cost often concentrates in validity, and validity is the component where artifact design has the clearest direct leverage. We do not claim to address novelty or significance verification. The burden of those judgments also grows as submission volume grows, and it is real---but it calls for responses of its own, and it is not where the artifact-level intervention developed here is most likely to pay off.

\subsection{Why Validity Verification Is Expensive in Prose}\label{sec:verification:prose}

Validity is not intrinsically expensive to verify. It is expensive to verify \emph{given the artifact form in which scientific arguments are currently delivered}. The dominant form is the prose paper, and prose has a specific structural property that makes validity verification costly: it serializes an argument graph into a reading order designed for narrative flow rather than for inspection.

When an author writes a paper, they hold something like a graph in mind: a set of claims, the evidence they have for each, the assumptions they are making, the dependencies among the claims, and the risks and limitations they have noticed. Writing the paper compresses that graph into a linear text optimized for clarity, persuasion, and conformity to a venue's expectations. The compression discards information. Implicit assumptions are not written down because they would be tedious to enumerate. Dependencies between claims are signaled by transitional phrases rather than recorded as edges. Limitations may be acknowledged in a paragraph and not propagated to the claims they qualify. That prose leaves the structure of an argument implicit is a long-standing observation in argumentation theory and semantic publishing, which have proposed explicit representations of claims, grounds, issues, and their support relations for exactly this reason~\citep{toulmin2003uses,kunz1970issues,groth2010anatomy,clark2014micropublications}.

A reviewer who wants to verify the paper must reverse this compression. They identify the central claims, which may be distributed across the abstract, introduction, results, and conclusion in slightly different forms; locate the evidence for each, in methods, figures, appendices, or prior work; infer implicit assumptions; check whether the asserted support relations actually hold, which often requires recomputing or rereading; and track risks and contradictions, which may be acknowledged in one place and silently violated in another. Each of these is a reconstruction of structure that existed in the author's head but was not recorded in the artifact. The author paid the structure cost once; or, increasingly, the author skipped it entirely, leaving an AI system to produce a serialization without the underlying graph.

This is also why AI verification of prose papers is harder than the surface symmetry between AI generation and AI verification might suggest. A long prose argument can exceed the context window or reasoning depth that current models can reliably handle for non-trivial scientific work, and the linearization removes precisely the local structure that an AI system could otherwise check piece by piece. The scientific claim-verification literature shows this at increasing scales (\Cref{app:benchmarks}): verification of atomic claims against abstracts becomes substantially harder in open-domain settings~\citep{wadden2020fact,wadden2022scifact}; validating dispersed claim--evidence links in full AI papers exposes precision--recall tradeoffs~\citep{javaji2025can}; manuscript-level error detection and paper-versus-reproduction-package checks report low precision, recall, and accuracy~\citep{Son2025WhenAC,hu2025repro}; and execution-grounded evaluation surfaces failures that narrative-only review misses~\citep{bai2026story}. Taken together, these benchmarks suggest that the bottleneck is not only model capability; it is the need to reconstruct from prose the claim--evidence and paper--code relations that the artifact did not make explicit. Better models will help, but the artifact form is itself the constraint. A verifier---human or AI---cannot easily check what the artifact has not recorded.

\subsection{What This Implies for Artifact Design}\label{sec:verification:lever}

The diagnosis points to a specific design lever. If a major verification cost is reconstruction of argument structure that the author had but did not record, the corresponding remedy is for the artifact to record that structure in the first place.

Two design desiderata follow. The first is to make the artifact \emph{structured rather than prose-first}: the argument graph is recorded explicitly as the primary representation, and prose, where it is needed, is generated from or attached to the graph rather than serving as the graph's only form. The second is to make the artifact \emph{decomposed rather than monolithic}: the graph is broken into shorter local arguments that can be inspected, doubted, or amended independently, rather than being packaged as a single linear narrative whose components cannot be addressed in isolation.

Together, these two desiderata are designed to redistribute the structure work from many downstream verifiers to one upstream author. The author pays an upfront cost---the cost of recording explicitly what would otherwise have been left implicit---with the aim that every downstream verifier pays less. This is what rebalancing the costs of generation and verification would look like at the artifact level. Section~\ref{sec:blueprints} develops what an artifact built around these two desiderata actually looks like.

%% file: sections/4-solution.tex
\section{Blueprints}\label{sec:blueprints}
Blueprints build on a long line of work arguing that scholarly knowledge should be represented as structured, machine-actionable claims rather than only as prose~\citep{shotton2009semantic,kuhn2017genuine,clark2014micropublications,groth2010anatomy}. The closest operational inspiration for our prototype is \texttt{leanblueprint}\footnote{\url{https://github.com/PatrickMassot/leanblueprint}} in formal theorem proving~\citep{de2015lean}, because it demonstrates how a long argument can be decomposed into locally inspectable graph components. This section describes the prototype only as far as needed to make the position concrete; \Cref{app:vocabulary,app:authoring,app:lint,app:interfaces,app:review-dialogue} give the full design.

\subsection{Background: \texttt{leanblueprint}}\label{sec:blueprints:lean}

\texttt{leanblueprint} is a tool in the Lean theorem-prover community~\citep{de2015lean} for planning, decomposing, and tracking the formalization of long proofs: a blueprint is a typed graph whose nodes are mathematical statements and whose edges record dependency, with a per-node flag for whether Lean has checked it. \Cref{app:leanblueprint} gives detail, including the Polynomial Freiman--Ruzsa formalization that used a blueprint to coordinate many contributors, non-experts included.

Two features instantiate the desiderata of Section~\ref{sec:verification:lever}. First, the artifact \emph{is} the structure: the graph records the proof's logical skeleton directly, and the prose attached to each node is supplementary, so a reviewer can inspect a single lemma and its dependencies without reading a paper. Second, the structure makes verification \emph{distributable}: contributors can work on individual lemmas in parallel, and the global proof status is a function of the per-node statuses. But \texttt{leanblueprint} works because Lean provides a formal proof obligation: a node is either machine-verified or it is not. General research has no such criterion, so \texttt{leanblueprint} cannot be transplanted directly. Our proposal adapts this kind of structured, decomposed argument graph to settings where verification is non-binary, where vocabularies vary by field, and where human and AI agents are both first-class users.

\subsection{The Design of Blueprints}\label{sec:blueprints:structure}

\paragraph{Data model.}\label{sec:blueprints:datamodel} A blueprint is a heterogeneous directed graph. Nodes and edges are both typed. Each node carries an \emph{ID}, \emph{type}, \emph{status}, \emph{label}, and an optional Markdown \emph{body} with mathematical notation; each edge carries an \emph{ID}, \emph{type}, and optional \emph{body} of the same form. Beyond the typing, no further structure is built into the data model: a blueprint is one JSON document, and can be inspected, edited, or transmitted as a single file. \Cref{fig:blueprint-example} shows a fragment of a real blueprint; \Cref{fig:blueprint-datamodel} in \Cref{app:vocabulary} illustrates every node and edge type.

\paragraph{Vocabulary.}\label{sec:blueprints:vocabulary} The built-in vocabulary comprises seven node types in four roles---\textsf{claim}, \textsf{evidence}, and \textsf{assumption} (the argument core); \textsf{definition} and \textsf{question} (construction handles for building the graph without asserting propositions); \textsf{risk} (a review channel that attaches a counter-argument to a specific claim, in place); and \textsf{synthesis} (a flexibility valve that carries prose when a joint argument does not fit on a single edge)---together with five edge types (\textsc{supports}, \textsc{expands}, \textsc{contradicts}, \textsc{addresses}, \textsc{relates}) and a short, type-specific status ladder for each node type. This lightweight core adapts Toulmin's claim--data--warrant structure~\citep{toulmin2003uses}, while the surrounding graph vocabulary draws on issue-based discourse models such as IBIS~\citep{kunz1970issues}; it is the built-in default and can be overridden per document or per workspace (\Cref{app:vocabulary}).

\subsection{Using Blueprints}\label{sec:blueprints:use}

\paragraph{Two interfaces over a shared document.}\label{sec:blueprints:use:interfaces} The prototype exposes two first-class interfaces over a shared blueprint document: a browser canvas, where a human user edits the graph as an interactive mindmap, and a command-line interface (CLI), through which an AI agent issues the same edits---we have used Claude Code and Codex as the primary test agents. Both interfaces read and write the same JSON file, and a development server propagates edits between them within roughly 200 milliseconds, so a researcher and their agent can edit the same artifact in parallel rather than coordinating through a chat window (\Cref{app:interfaces}).

\paragraph{Authoring.}\label{sec:blueprints:use:authoring} The authoring workflow follows \texttt{leanblueprint}'s in spirit. An author builds the argument core from \textsf{claim}, \textsf{evidence}, and \textsf{assumption} nodes connected by \textsc{supports} edges; scaffolds it with \textsf{definition} and \textsf{question} nodes; and tracks progress through node statuses, advancing them bottom-up so that a \textsf{claim} is not marked \texttt{supported} before its supporters have matured (\Cref{app:authoring}). Ideally a blueprint begins before the paper: just as a \texttt{leanblueprint} precedes formalization, it can start as a planning artifact of draft claims and open questions, evolve as the research proceeds, and eventually serve as the structured basis from which the paper is written---the paper becoming a narrative rendering of the matured graph. We expect initial adoption to be largely retrospective.

\begin{figure}[t]
\centering
\includegraphics[width=0.9\textwidth]{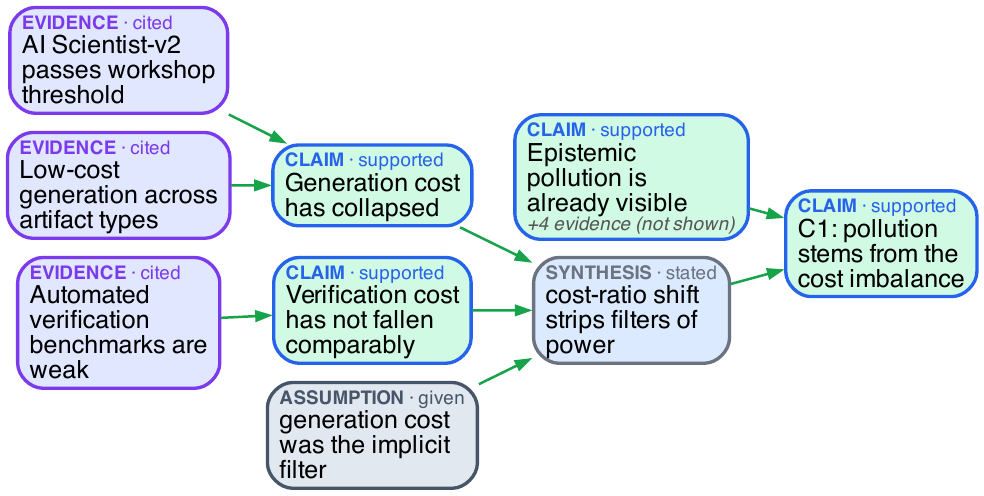}\\[3pt]
\includegraphics[width=0.88\textwidth]{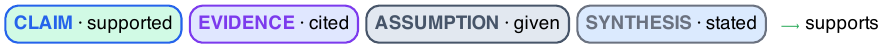}
\caption{A fragment of the blueprint of this paper as an illustration. The three premises establish the conclusion only jointly, so they enter a \textsf{synthesis} node instead of supporting the claim through independent edges. The full blueprint is in the accompanying GitHub repository (\blueprintRepoUrl).}
\label{fig:blueprint-example}
\end{figure}

\paragraph{Synthesis as a flexibility valve.} Some arguments rest on premises that interact, and forcing them onto independent edges would mislead. This paper's own argument supplies an example (\Cref{fig:blueprint-example}). The claim that existing filters lost their discriminating power rests jointly on three premises: that generation cost historically acted as an implicit filter, that generation cost has collapsed, and that verification cost has not fallen comparably. None of the three independently supports the conclusion, because the argument concerns their \emph{ratio}; three independent \textsc{supports} edges would misstate it as three sufficient routes. A \textsf{synthesis} node represents the joint inference. The valve is also the type most easily misused: its prose body can hide structure that should have been decomposed, so lint flags suspicious synthesis patterns, and a blueprint that leans heavily on \textsf{synthesis} signals that decomposition has not been pushed hard enough.

\paragraph{Reviewing.}\label{sec:blueprints:use:reviewing} A blueprint is reviewed---by the author, a human reviewer, or an AI reviewer---through structural inspection of local subgraphs and through a dialogue the vocabulary supports: statuses record the author's current positions, and a reviewer contests a position structurally by attaching a \textsf{risk} node that the author then \textsc{addresses} or accommodates by revising statuses (\Cref{app:review-dialogue}). The prototype also ships around a dozen structural lint rules---for instance, flagging a \textsf{claim} marked \texttt{supported} with no incoming \textsc{supports} edges, which is structurally an assumption rather than a claim. The lint rules serve as warnings instead of blocks: lint can flag patterns but cannot adjudicate content, so the reviewer remains the arbiter (\Cref{app:lint}).

\paragraph{Distributed and local review.} The intended payoff of the structure is that review can become more local. A reviewer can pick a \textsf{claim} and inspect its supporters, evaluate the \textsf{evidence} bodies and \textsf{assumption} justifications attached to that local subgraph, and form a judgment on that subgraph without first reconstructing the entire argument from prose. Different subgraphs can in principle be reviewed by different reviewers in parallel: an AI critic can enumerate counter-arguments as new \textsf{risk} nodes on one matured-claim subgraph, while an expert verifies the citation behind a single \textsf{evidence} node. This is the property the rebalancing argument of Section~\ref{sec:verification:lever} ultimately rests on: an artifact that records structure is designed to make verification distributable.

%% file: sections/5-related.tex
\section{Related Work}\label{sec:related}

\paragraph{Structured scholarly artifacts and semantic publishing.}
Blueprints sit within a broader semantic-publishing tradition that argues scientific communication should expose machine-actionable scholarly structure rather than leave knowledge locked inside prose~\citep{shotton2009semantic,kuhn2017genuine}. Nanopublications proposed RDF-based packages for core scientific statements and their provenance, while micropublications developed a richer semantic model for claims, evidence, arguments, and annotations~\citep{groth2010anatomy,clark2014micropublications}; later work has explored nanopublication-based reviewing workflows~\citep{bucur2023nanopublication}. The Open Research Knowledge Graph represents research contributions as structured scholarly knowledge~\citep{jaradeh2019orkg}, CiTO makes scholarly relations such as citation intent explicit~\citep{shotton2010cito}, and Octopus decomposes research into typed, independently publishable components~\citep{freeman2023octopus}. Blueprints therefore do not claim novelty in representing claims, evidence, provenance, or scholarly relations as structured data. They build on this lineage but focus on the AI-era problem of lowering downstream validity-verification cost through a lightweight, author-facing artifact with local review statuses, soft structural lint, and shared human/agent interfaces. This lineage also sharpens a teleological question: if prose is suboptimal for verification, why has it dominated, and why would a structured alternative take hold now? Prose became dominant under a cost regime in which generating a credible artifact was the scarce, costly act and human reading was the only interface; earlier structured-communication efforts stalled largely because the authoring cost fell on authors with no offsetting payoff under that regime. AI-era science inverts both terms: verification is now the scarce act, and AI agents can bear much of the structuring cost (\Cref{sec:blueprints:use:authoring}).

\paragraph{Argumentation models and scientific claim verification.}
The blueprint vocabulary is not intended as a new theory of argument. It is a pragmatic vocabulary for research artifacts, drawing from long-standing models of claims, grounds, rebuttals, issues, positions, and arguments~\citep{toulmin2003uses,kunz1970issues}. Scientific claim-verification, manuscript error-detection, reproducibility, and execution-grounded benchmarks (\Cref{sec:verification:prose}, \Cref{app:benchmarks}) mostly operate downstream of already-written artifacts, recovering or assessing support relations, claim--evidence links, or narrative--execution consistency after the fact. Blueprints instead make a project's support structure part of the artifact before verification begins.

\paragraph{Provenance, infrastructure, and agent-native artifacts.}
Other responses to AI-era scientific risk improve provenance and infrastructure~\citep{earp2025llm,Ghosh2025AnII,pei2025open}. A concurrent proposal for agent-native research artifacts makes a closely related claim: papers discard structure that AI agents need in order to inspect, reproduce, and extend research~\citep{liu2026last}. Blueprints are complementary but narrower: rather than recording attribution and lineage or packaging research for execution, they target a lightweight layer for validity verification, recording how claims, evidence, assumptions, risks, and definitions support one another (\Cref{app:related}).

%% file: sections/6-alternatives.tex
\section{Alternative Views and Limitations}\label{sec:alternatives}

We consider positions opposed to ours, and limitations of the specific proposal, in turn.

\paragraph{Validity is not the binding constraint.} One may object that rebalancing through validity misses the real burden: novelty and significance judgments also scale with submission volume, and validity can only truly be checked by reproducing the work. We agree that a validity-focused intervention mitigates rather than resolves the imbalance (\Cref{sec:verification:validity}), and that empirical reproduction needs complementary treatment (\Cref{sec:verification:components}). Our claim is narrower: argument validity is a significant and currently under-addressed cost. Many documented failure modes of AI-generated manuscripts, such as hallucinated citations and fabricated papers whose plausible claims lack grounding~\citep{naddaf2026hallucinated,haider2024gpt}, are failures of the support relations between claims and evidence rather than of writing style or empirical execution, and they remain hard to catch today~\citep{Son2025WhenAC}.

\paragraph{Alternative interventions.} The generation-verification imbalance might be addressed without redesigning scientific artifacts: by improving AI verifiers and reviewers, by expanding human reviewer capacity and incentives, or by introducing better peer-review mechanisms. We view these directions as valuable and complementary rather than competing. Section~\ref{sec:pollution} discusses why better AI alone may not be enough; interventions on reviewer capacity or incentives alone, in turn, do not change the underlying cost economics---they add capacity to the expensive side of the ratio without making any artifact cheaper to check. There is also a dependency running in the other direction: ecosystem-level mechanisms one might eventually want, such as claim-level citation, subgraph-level reviewer assignment, or cross-paper contradiction detection, presuppose artifacts that record structure.

\paragraph{Technical counterarguments.} Several counterarguments may target the blueprint design itself. \emph{Expressiveness loss}: graphs may not capture the semantic richness and nuance of prose. The design partially mitigates this through Markdown bodies on every node and edge and through \textsf{synthesis} nodes as a flexibility valve; more fundamentally, blueprints complement rather than replace papers, so what the graph does not express deliberately remains in the paper. \emph{Poor decompositions}: a graph can be badly factored, with an argument's real difficulty hidden inside one large or opaque node. This is a genuine limitation. Structural lint flags symptoms such as matured claims without grounded support, but it cannot determine whether a decomposition is substantively adequate. The reviewer remains the arbiter. Semantic checking tools that flag unusually broad or insufficiently verifiable nodes are a possible future mitigation, with final judgment still human. \emph{Gaming}: an adversary can generate a plausible-looking blueprint just as they can generate plausible prose. Blueprints do not eliminate this risk. Their advantage is narrower: commitments become explicit and locally inspectable, so a reviewer can challenge an individual support relation, identify circular dependencies, or check whether a cited evidence node actually supports the claim its edge asserts. Fabrication is not prevented, but some forms of inconsistency become easier to detect locally.

\paragraph{Systemic risks.} Two risks operate at the level of the scientific system rather than the artifact. \emph{Restricting plurality of interpretation}: any canonical structured representation risks narrowing how an argument can be interpreted. The current design mitigates but does not eliminate this risk---a blueprint records the author's asserted argument rather than imposing a field-level ontology, statuses are contestable assertions that reviewers can challenge through \textsf{risk} nodes and status revisions, and vocabularies can be overridden per document---but the possibility that a shared representation privileges one reading over others remains an open problem. \emph{Partial cost redistribution}: if blueprints shift only part of the verification cost, venues might raise expectations for authors without meaningfully reducing verifier burden, and the community could fragment into blueprint- and prose-centered cultures with additional coordination costs. These are adoption-dynamics questions that cannot be resolved from the artifact design alone; they belong on the empirical agenda of \Cref{sec:discussion}.

\paragraph{Disciplinary scope.} The blueprint vocabulary is most directly applicable to research whose central contributions can be organized around claims, evidence, assumptions, and support relations. Interpretive, ethnographic, and other research traditions may fit this representation poorly, and the extensible vocabulary of \Cref{sec:blueprints:vocabulary} is a hypothesis about adaptability, not evidence that the framework generalizes across disciplines. We state this as an explicit scope condition rather than treat science as a monolithic enterprise: the position argued here concerns the artifact forms of claim-and-evidence-structured research, and its applicability beyond them remains to be established.

%% file: sections/7-discussion.tex
\section{Discussion}\label{sec:discussion}

\paragraph{Engineering of science as a research program.} The argument presented here exemplifies a broader principle: scientific artifacts and infrastructure should be designed intentionally and evaluated systematically against criteria such as verification costs, error rates, ease of adoption, and resilience against adversarial or low-effort use. This principle is already implicitly recognized across open science, reproducibility, semantic publishing, and scholarly infrastructure. The cost imbalance introduced by AI makes one design question urgent: which elements of scientific artifacts require redesign now that plausible generation is inexpensive, yet rigorous verification remains resource-intensive?

\paragraph{Empirical validation of blueprints.} This paper focuses on staking out a position and proposing a preliminary prototype design. The claim that recording argument structure upstream can reduce the reconstruction burden paid by downstream verifiers is a hypothesis that has yet to be fully tested. We see four natural starting points for empirical validation: studies of blueprint authoring overhead and verification time on small case-study projects, comparison of reviewer agreement and error-detection rates between prose papers and corresponding blueprints, validation of vocabulary extensions in disciplines outside our default, and studies of the adoption dynamics raised in \Cref{sec:alternatives}---whether blueprint requirements shift costs onto authors without relieving verifiers.

\paragraph{Beyond papers, and beyond verification.} Research proposals may be an especially promising early application: a proposal is an argument about work not yet completed, whose claims rest on assumptions, preliminary evidence, and open questions that \textsf{question} nodes and \textsf{evidence} statuses can make explicit. Blueprints may also matter for generation: just as test suites make code-generation agents more useful by giving them local targets and structured feedback, explicit claims and support relations give an AI agent something to extend, check, and push against. In this sense blueprints are a candidate harness for AI-assisted science---they make verification more local now, and may make generation more disciplined later (\Cref{app:discussion}).

%% file: sections/8-conclusion.tex
\section{Conclusion}\label{sec:conclusion}

AI has made plausible scientific artifacts cheap to generate while leaving reliable verification comparatively expensive. This is not only a model-capability problem, but an infrastructure problem: scientific systems built around papers, peer review, and citation were calibrated to a world in which plausibility itself was expensive to produce.

We have argued for one artifact-level response: blueprints, structured and decomposed argument graphs that make support relations inspectable before a verifier has to reconstruct them from prose. Papers remain the narrative interface of science; blueprints are proposed as a verification interface. The broader argument is that AI-era scientific artifacts should be designed for verification as deliberately as they are designed for communication, so the infrastructure that admits scientific claims can keep pace with the systems that generate them.

%% file: sections/appendix.tex
\section{Background on \texttt{leanblueprint}}\label{app:leanblueprint}

This appendix expands the background summarized in \Cref{sec:blueprints:lean}. \texttt{leanblueprint} is a tool used in the Lean theorem prover community~\citep{de2015lean} to plan, decompose, and track formalization of long mathematical proofs. A ``blueprint'' in \texttt{leanblueprint} is a typed graph: nodes are mathematical statements (definitions, lemmas, theorems), and edges record dependency---which lemmas a theorem rests on, or which definitions a lemma uses. Each node carries a status flag indicating whether it has been formally checked by Lean. The canonical demonstration of the use of \texttt{leanblueprint} is the Lean formalization of the proof of the Polynomial Freiman--Ruzsa conjecture, where Tao and collaborators used a \texttt{leanblueprint} to coordinate a distributed effort across many contributors, including non-experts who could pick up local lemmas without holding the full proof in mind\footnote{\href{https://terrytao.wordpress.com/2023/11/18/formalizing-the-proof-of-pfr-in-lean4-using-blueprint-a-short-tour/}{Tao's PFR Lean blueprint tour}.}.

\section{Blueprint Vocabulary}\label{app:vocabulary}

\begin{figure}[!htbp]
\centering
\includegraphics[width=0.62\textwidth]{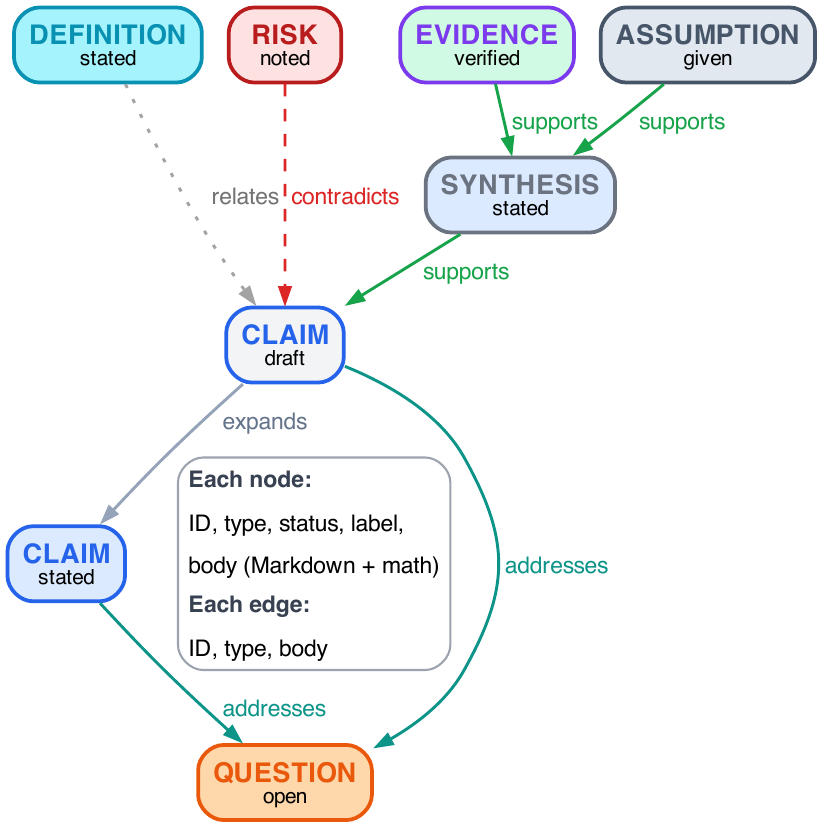}
\caption{The blueprint data model: a heterogeneous graph with typed nodes and typed edges. Each node shows its type (border color) and status (fill), and each edge is labeled with its type. Every node carries an ID, type, status, label, and an optional Markdown body with mathematical notation; every edge carries an ID, type, and optional body.}
\label{fig:blueprint-datamodel}
\end{figure}

This appendix gives the full built-in vocabulary summarized in \Cref{sec:blueprints:vocabulary}; \Cref{fig:blueprint-datamodel} illustrates it. The built-in vocabulary consists of seven node types organized into four roles, five edge types, and a small collection of node-type-specific status ladders.

\paragraph{Node types: argument core.} Three node types carry the load-bearing content of an argument. This lightweight vocabulary adapts Toulmin's claim--data--warrant structure~\citep{toulmin2003uses}, while the surrounding graph vocabulary draws on issue-based discourse models such as IBIS~\citep{kunz1970issues}. A \textsf{claim} is a proposition the author proposes to defend. An \textsf{evidence} is a proposition reporting an observation, a measurement, or a citation that bears on a claim. An \textsf{assumption} is a proposition taken as given.

\paragraph{Node types: construction handles.} Two node types help the author build the graph without asserting propositions. A \textsf{definition} pins a term so that the rest of the argument can refer to it precisely. A \textsf{question} marks an unresolved direction.

\paragraph{Node types: review channel.} A \textsf{risk} node attaches a counter-argument or suspicion to a specific claim, in place.

\paragraph{Node types: flexibility valve.} A \textsf{synthesis} node carries prose when a joint argument across several premises does not fit on a single edge body.

\paragraph{Edge types.} Five edge types connect nodes: \textsc{supports} (an \textsf{evidence}, \textsf{assumption}, or supporting \textsf{claim} backs a downstream \textsf{claim}), \textsc{expands} (a child node elaborates its parent), \textsc{contradicts} (a tension between two nodes), \textsc{addresses} (a node responds to a \textsf{risk} or \textsf{question}), and \textsc{relates} (a non-committal association).

\paragraph{Statuses.} Each node type carries a status from a short type-specific ladder. A \textsf{claim} moves through \texttt{draft}, \texttt{stated}, and \texttt{supported}. An \textsf{evidence} moves through \texttt{missing}, \texttt{cited}, and \texttt{verified}. A \textsf{definition} is \texttt{draft} or \texttt{stated}. A \textsf{question} is \texttt{open} or \texttt{answered}. An \textsf{assumption} is \texttt{given} or \texttt{questioned}. A \textsf{risk} is \texttt{noted}, \texttt{addressed}, or \texttt{accepted}. A \textsf{synthesis} is \texttt{draft} or \texttt{stated}.

\paragraph{Extensibility.} The vocabulary above is the built-in default. The blueprint format allows a per-document override (encoded in the document itself), a per-workspace override (a single configuration file applying to every document in the workspace), and the built-in default.

\section{Authoring a Blueprint}\label{app:authoring}

This appendix expands the authoring workflow summarized in \Cref{sec:blueprints:use:authoring}. The authoring workflow is similar in spirit to \texttt{leanblueprint}'s, with adaptations forced by the move from formal mathematics to general research. An author builds the graph by drafting propositional content, scaffolding it with construction handles, tracking progress through statuses, and reaching for synthesis only as a flexibility valve. Authoring can also be partly delegated to an AI agent operating through the CLI.

\paragraph{Drafting the argument core.} The propositional content of an argument is built up using \textsf{claim}, \textsf{evidence}, and \textsf{assumption} nodes connected by \textsc{supports} edges. An author typically starts with a small number of central \textsf{claim} nodes, then either decomposes each into supporting sub-claims or grounds it in \textsf{evidence} and \textsf{assumption} nodes. Edges may carry their own bodies that explain why the source backs the target. A useful authoring rule of thumb: if the supporting premises do not interact---each supports the conclusion independently---use separate edges with bodies; if they interact and the joint argument must be considered as a whole, use a \textsf{synthesis} node.

\paragraph{Scaffolding with construction handles.} Two non-propositional node types help the author build the graph. A \textsf{definition} pins a term so that the rest of the argument can refer to it precisely; pinning a definition early lets later claims use the term without re-explaining it each time. A \textsf{question} marks an unresolved direction during authoring---a thread the author has not yet decided how to resolve, or a sub-problem to come back to. The same node type doubles as a future-work flag in a finished blueprint, which means questions don't need to be deleted before sharing; they signal what's still open.

\paragraph{Tracking progress through statuses.} Each node carries a status from a short type-specific ladder. During authoring, statuses act as a progress tracker: a \textsf{claim} is \texttt{draft} while sketched, \texttt{stated} once precisely formulated, and \texttt{supported} when the author asserts it is settled by its incoming support edges. An \textsf{evidence} moves through \texttt{missing} (node added but without content), \texttt{cited} (linked to the source), and \texttt{verified} (independently checked). Authoring discipline is to advance statuses bottom-up: an author should not mark a \textsf{claim} \texttt{supported} before its supporting \textsf{evidence} and supporting \textsf{claim}s have themselves been matured. Doing so triggers structural warnings (\Cref{app:lint}), which is the lint helping the author notice they have asserted completeness that the structure does not yet warrant.

\paragraph{Synthesis as a flexibility valve.} A \textsf{synthesis} node is used when several supporters establish a downstream claim only by being considered together---an argument that does not cleanly distribute across separate edges (see the example in \Cref{fig:blueprint-example}). But \textsf{synthesis} is also the type most easily misused. Its body hides structure inside prose, which is the opposite of what the typed graph is for, and a blueprint that leans heavily on \textsf{synthesis} may signal that decomposition has not been pushed hard enough.\footnote{\texttt{leanblueprint} faces a similar tension: a poorly decomposed proof can hide all its difficulty inside one large lemma.} The artifact does not adjudicate. Lint warns about specific misuse patterns---empty synthesis nodes, single-supporter synthesis---but the deeper question of whether a synthesis is genuinely irreducible is left to the author.

\paragraph{Co-authoring with an AI agent.} Because the CLI exposes the same operations as the canvas, an AI agent can be given partial authoring tasks: populating evidence bodies from cited sources, suggesting decompositions of a sketched claim, or extending a subgraph along a direction the author has indicated. Coordination between human and agent currently happens at the prompt level---the author tells the agent which subgraph to work on---rather than through ownership primitives in the artifact itself.

\section{Structural Lint Rules}\label{app:lint}

This appendix details the structural lint rules summarized in \Cref{sec:blueprints:use:reviewing}. The artifact ships with around a dozen structural rules grouped along three axes. All rules are warnings, not blocks: the artifact does not refuse to save or share a blueprint that has unresolved warnings, because lint can flag patterns but cannot adjudicate content. The reviewer is the arbiter; the lint surfaces what the reviewer should look at.

The rules of the first axis check the \emph{soundness of the support graph}. A \textsc{supports} edge running the wrong way (for instance from a claim into its own evidence) is almost always a typo or a confused edge. A \textsc{supports} edge into a matured claim from a node that is not itself grounded indicates the chain of support has a hole. The most informative rule on this axis flags a \textsf{claim} marked \texttt{supported} with no incoming \textsc{supports} edges at all. Structurally, such a node is an assumption, not a claim: the author is asserting a proposition without offering anything to back it. The lint does not insist the author convert it to an \textsf{assumption}---if the proposition really should be backed, the author needs to add the supports---but the warning makes the type-vocabulary distinction operational. The difference between a claim and an assumption is exactly what the support graph records.

The rules of the second axis check \emph{argument structure}. Directed cycles in the \textsc{supports} subgraph are flagged because circular support is not support. Single-supporter \textsf{synthesis} nodes are flagged because synthesis is for joint support, and a synthesis with one supporter is using the wrong type. The rules of the third axis check \emph{vocabulary integrity}: nodes or edges with type or status identifiers not in the active registry, or with empty labels, are flagged because they cannot be reasoned about reliably.

\section{The Two Blueprint Interfaces}\label{app:interfaces}

\begin{figure}[!htbp]
\centering
\includegraphics[width=\textwidth]{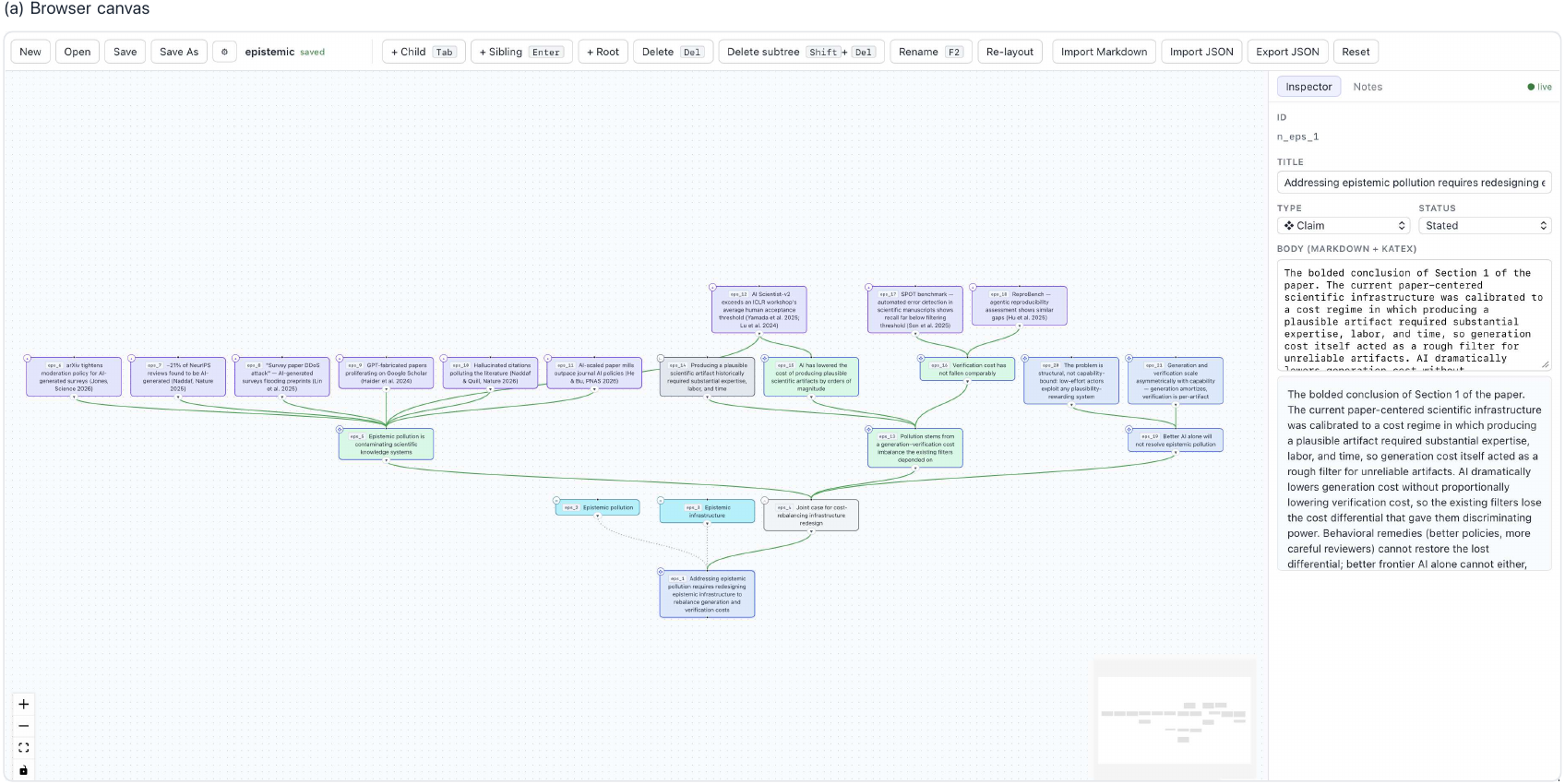}\\[6pt]
\includegraphics[width=0.7\textwidth]{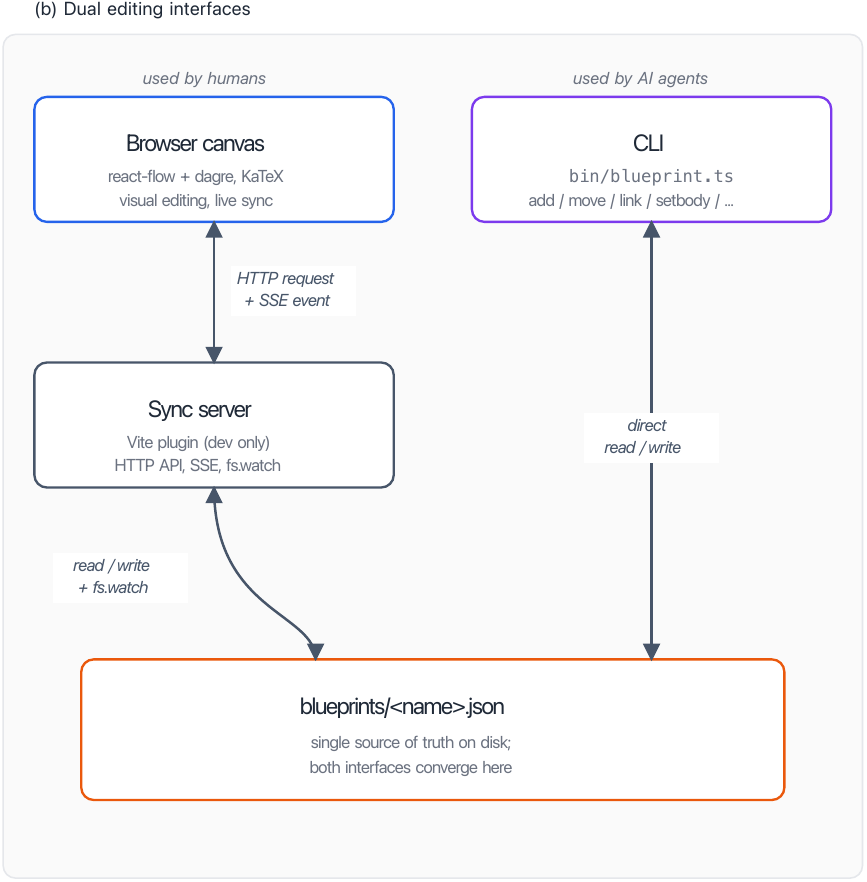}
\caption{Dual interfaces over a shared blueprint document. Top (a): the browser canvas, showing an early blueprint of this paper as an interactive typed mindmap, with the inspector panel on the right. Bottom (b): the architecture of the two interfaces, which converge on a single JSON file on disk.}
\label{fig:blueprint-interfaces}
\end{figure}

This appendix expands the dual-interface description summarized in \Cref{sec:blueprints:use:interfaces}. The prototype exposes two first-class interfaces, respectively for human users and AI agents, over a shared blueprint document (\Cref{fig:blueprint-interfaces}).

\paragraph{Browser canvas for human users.} The blueprint renders as an interactive mindmap on a browser canvas, where a human user can intuitively edit the graph; an inspector panel shows and edits the selected node's or edge's fields.

\paragraph{Command-line interface (CLI) for AI agents.} The edits on a blueprint can also be done through a set of CLI commands. The commands are designed to be issued by an AI agent rather than typed by a human; we have used Claude Code and Codex as the primary test agents, each given a small skill description that explains when and how to use the available commands.

\paragraph{Live synchronization.} Both interfaces read and write the same JSON file storing the blueprint. A development server watches the workspace directory and propagates edits between the two within roughly 200 milliseconds: when the agent issues a CLI command that mutates the file, the human's canvas updates live; when the human edits a node body in the canvas, the agent's next read reflects the change. The shared-substrate property is what makes a researcher and their agent able to edit the same artifact in parallel rather than coordinating through a chat window.

\section{The Blueprint as a Review Dialogue}\label{app:review-dialogue}

This appendix develops the review-dialogue view summarized in \Cref{sec:blueprints:use:reviewing}. The blueprint represents an argument as it is being made, not as it has been proved. Statuses are the author's positions: a \texttt{supported} \textsf{claim} is one the author asserts is settled by its support graph. A reviewer who challenges that assertion does so structurally, by attaching a \textsf{risk} node to the \textsf{claim}. The author responds by adding nodes that \textsc{addresses} the \textsf{risk}, by revising the support graph, or by moving the \textsf{claim}'s status back to \texttt{stated} while the contestation is unresolved. Statuses are revisable: their point is not to certify but to record the author's current position, so the review dialogue has somewhere concrete to start.

This is what \texttt{leanblueprint} does not need to do: in formal mathematics the kernel arbitrates, so a proved status is a fact, not a claim. In general research no kernel exists, and the artifact has to make explicit that statuses are author assertions, structurally tracked but contestable.

\section{Scientific Claim-Verification Benchmarks}\label{app:benchmarks}

This appendix expands the benchmark evidence cited in \Cref{sec:verification:prose}, where it is argued that verifying validity in prose is expensive for AI systems as well as humans. The scientific claim-verification literature makes this point at increasing scales. SciFact formulates verification as finding abstracts that support or refute an atomic scientific claim and selecting rationale sentences; SciFact-Open shows that the same task becomes substantially harder when evidence must be found in an open-domain corpus and when evidence only partially matches a claim's scope~\citep{wadden2020fact,wadden2022scifact}. CLAIM-Bench moves the problem into full AI papers, where models must extract claims, locate dispersed evidence spans, and validate the links between them, exposing precision--recall and computational-cost tradeoffs~\citep{javaji2025can}. At the manuscript and reproducibility levels the gap is larger still: SPOT reports low precision and recall for significant errors in published manuscripts, while REPRO-Bench reports low accuracy when agents must compare a paper PDF with its reproduction package~\citep{Son2025WhenAC,hu2025repro}. Execution-grounded evaluation systems make the same point constructively: checking code, data, and execution traces alongside the paper surfaces failures that narrative-only review misses~\citep{bai2026story}.

\section{Evidence of Epistemic Pollution}\label{app:pollution-evidence}

This appendix collects the evidence on peer review and adjacent infrastructure summarized in \Cref{sec:pollution}. A study in \emph{Nature} reported that 21\% of manuscript reviews for a major AI conference were found to be generated by AI~\citep{naddaf2025major}, raising the prospect of a feedback loop in which AI-generated submissions are evaluated by AI-generated reviews~\citep{kim2025position,wei2025ai}. More broadly, LLMs are now being tested as part of the review pipeline itself, including at scale and in randomized peer-review interventions~\citep{liang2024can,thakkar2026large}. Adjacent infrastructure is similarly affected. Large-scale evidence suggests that journal AI-use policies have not produced reliable disclosure or restrained AI-assisted writing~\citep{he2026academic}. Paper-mill operations are also scaling with generative AI, and documented cases of AI-generated articles falsely attributed to real researchers show that authorship metadata can itself be corrupted~\citep{spinellis2025false}. The authors of AI Scientist-v2 report that an end-to-end autonomous research system could formulate hypotheses, run experiments, and author a manuscript that exceeded an ICLR workshop's average human acceptance threshold for a fully autonomous submission~\citep{Yamada2025TheAS,Lu2024TheAS}.

\section{Extended Related Work}\label{app:related}

This appendix expands the related work in \Cref{sec:related}.

\paragraph{AI-era production pressure and peer review.}
Recent empirical work suggests that LLMs are already affecting scientific writing at scale and may be changing scientific production patterns~\citep{liang2025quantifying,kusumegi2025scientific}. Automated survey-writing systems show that literature synthesis itself can be generated cheaply~\citep{wang2024autosurvey}, and AI-generated survey volume has been framed as a denial-of-service attack on research attention~\citep{lin2025stop}. Related failure modes include GPT-fabricated papers, false authorship, and hallucinated citations~\citep{haider2024gpt,spinellis2025false,naddaf2026hallucinated,sakai2026hallucitation}. Peer review is also being reshaped: AI-generated reviews are already visible in major conferences~\citep{naddaf2025major}, while LLM feedback has been evaluated at scale and in randomized peer-review interventions~\citep{liang2024can,thakkar2026large}. These lines of work motivate the same bottleneck: as plausible artifacts and reviews become cheaper to produce, the value of making verification structurally easier increases.

\paragraph{Provenance, infrastructure, and agent-native artifacts.}
Another response to AI-era scientific risk is to improve provenance and infrastructure. Recent work frames LLM-assisted scholarly writing as a provenance problem~\citep{earp2025llm}, while broader proposals call for AI-era scientific infrastructure and renewed open-science systems~\citep{Ghosh2025AnII,pei2025open}. A very recent, concurrent proposal for agent-native research artifacts makes a closely related claim: papers discard structure that AI agents need in order to inspect, reproduce, and extend research~\citep{liu2026last}. Blueprints are complementary but narrower. Provenance systems record attribution, lineage, and production history; agent-native artifacts package research for execution, reproduction, and extension; blueprints target a lighter-weight layer for validity verification, recording how claims, evidence, assumptions, risks, and definitions support one another. Their purpose is to make the verification interface of a research project explicit.

\section{Extended Discussion}\label{app:discussion}

This appendix expands the forward-looking directions noted in \Cref{sec:discussion}.

\paragraph{Beyond papers: research proposals.} Papers are not the only artifacts whose verification cost matters. Research proposals may be an especially promising early application of the same design. A proposal is an argument about work not yet completed: claims about expected contributions rest on assumptions, preliminary evidence, methodological plans, and open questions, with no completed results to anchor a quick assessment---and grant review shares the per-verifier reconstruction structure of \Cref{sec:verification:components}. The blueprint vocabulary maps naturally onto this setting: \textsf{question} nodes can represent unresolved design decisions, and \textsf{evidence} statuses can distinguish established support from conjecture or planned validation.

\paragraph{Blueprints as harnesses for AI-assisted science.} Blueprints may matter not only for verification, but eventually for generation. In software, test suites do not merely check code after it is written; they also make code-generation agents more useful by giving them local targets, structured feedback, and a way to iterate without holding the entire system in context. Blueprints could play a similar role for scientific arguments. Because claims, evidence, assumptions, risks, and support relations are explicit, an AI agent can be asked not only to inspect a finished argument, but also to extend a subgraph, propose missing evidence, draft a synthesis node, identify unsupported claims, or generate alternative decompositions against the current structure. The resulting additions would still be provisional and subject to human judgment, but the artifact gives generation something to push against. In this sense, blueprints are a candidate harness for AI-assisted science: they make verification more local now, and may make generation more disciplined later.